\documentclass[journal]{IEEEtran}
\usepackage{graphicx}
\usepackage{color}
\usepackage{cite}
\usepackage{amsmath,amssymb}
\usepackage{threeparttable}

\usepackage{pgfplots}
\pgfplotsset{compat=1.9}
\usepackage{tabularx,booktabs}
\usepackage{wasysym}
\usepackage[bookmarks=false]{hyperref} 
\usepackage{float}
\newcolumntype{C}{>{\centering\arraybackslash}X} 
\usepackage{siunitx}
\usepackage{caption}

\usepackage{tikz}
\usepackage{tikzscale}
\usepackage{derivative}

\usepackage{dblfloatfix}
\usepackage{color}

\begin{document}

\title{Impact of E-plane Misalignment on THz Diagonal Horn Antennas}

\author{

Divya~Jayasankar,~\IEEEmembership{Graduate student~member,~IEEE},
Andre Koj, 
Jeffrey~Hesler,~\IEEEmembership{Fellow,~IEEE}, 
and Jan~Stake,~\IEEEmembership{Senior Member,~IEEE}

\thanks{\noindent Manuscript received June 18 2024; revised July xxxx; accepted Aug xxxx. Date of publication Sep xxxx; date of current version Aug xxxx. This work was supported in part by the Swedish Research Council (Vetenskapsrådet) under grant 2020-06187. This work is conducted within the Advanced Digitalization program at the WiTECH Centre in Cryter+ project financed by VINNOVA, Chalmers University of Technology, AAC Clydespace, Low Noise Factory (LNF), Research Institutes of Sweden (RISE) and Virginia Diodes, Inc. (VDI). Miss. Jayasankar's PhD project is supported by the Swedish Foundation for Strategic Research (SSF) project number FID17-0040 and the Swedish Innovation Agency (Vinnova) project 2022-02967. \textit{(Corresponding author: Divya Jayasankar e-mail: divyaj@chalmers.se)} }
\thanks{D. Jayasankar is with the Terahertz and Millimetre Wave Laboratory, Department of Microtechnology and Nanoscience (MC2), Chalmers University of Technology, SE-412 96 Gothenburg, Sweden, and also with the Research Institutes of Sweden (RISE), SE-504 62 Borås, Sweden.}
\thanks{J. Hesler is with the Virginia Diodes Inc. (VDI), Charlottesville, VA 22902 USA, and also with the Terahertz and Millimetre Wave Laboratory, Department of Microtechnology and Nanoscience (MC2), Chalmers University of Technology, SE-412 96 Gothenburg, Sweden.}
\thanks{A. Koj was with the Terahertz and Millimetre Wave Laboratory. He is now with the Quantum Technology group, Department of Microtechnology and Nanoscience (MC2), Chalmers University of Technology, SE-412 96 Gothenburg, Sweden.}
\thanks{J. Stake is with the Terahertz and Millimetre Wave Laboratory, Department of Microtechnology and Nanoscience (MC2), Chalmers University of Technology, SE-412 96 Gothenburg, Sweden.}
\thanks{Color versions of one or more figures in this article are available online at http://ieeexplore.ieee.org. Digital Object Identifier 10.1109/TTHZ.2024.xx}}

\markboth{To be submitted to IEEE TRANSACTIONS ON TERAHERTZ SCIENCE AND TECHNOLOGY, 2024}%
{Jayasankar \MakeLowercase{\textit{et al.}}}

\markboth{To be submitted to IEEE TRANSACTIONS ON TERAHERTZ SCIENCE AND TECHNOLOGY, 2024}%
{Jayasankar \MakeLowercase{\textit{et al.}}}

\maketitle

\begin{abstract}
A key challenge in developing terahertz front-ends is achieving high coupling efficiency between the waveguide feed horn and the optical beam. In this paper, we have quantified the alignment requirements for the widely used E-plane split diagonal horn antenna through theoretical analysis, electromagnetic simulation, and experimental validation within the 325-500 GHz frequency range. The results from our analytical models, simulations, and measurements are consistent and shows good agreement. They reveal that even minor geometric asymmetries can cause significant increases in fractional power radiated to the cross-polar component due to amplitude and phase imbalances in the TE${10}$ and TE${01}$ modes. Furthermore, a misalignment of approximately
8\% of the wavelength was observed to result in a 3-dB degradation in the optical coupling to a Gaussian beam (Gaussicity) in middle of the waveguide band. These findings highlight the critical importance of precise alignment and feed horn machining for the successful implementation of terahertz front-end systems.
\end{abstract}



\IEEEpeerreviewmaketitle

\section{Introduction}

\IEEEPARstart{F}{eed} horns \cite{Barrow1939} play a pivotal role in various terahertz quasi-optical applications such as astronomy \cite{Farrah2019}, remote sensing \cite{Jeanne2023, Hammar2018}, imaging \cite{Zac2018}, material characterization \cite{Elena2022}, and spectroscopy \cite{Voigt23}. These applications often use horns integrated with detectors to maximize the signal coupling to non-linear elements like diodes,\cite{Gaidis, Reck23} or to radiate to free-space \cite{Ellison2019}. For astronomical receivers, corrugated feed horns an ideal candidate \cite{Wylde1984}. Electroforming has effectively produced corrugated horns with high Gaussicity (98\%), but achieving uniform $\lambda$/4 corrugations and coating at sub-millimeter wavelengths is challenging and expensive \cite{buchel2015}. Alternative horn profiles, such as Pickett-Potter horns \cite{potter1963}, conical horns \cite{King1950}, and smooth-walled spline horns \cite{granet2004}, offer easier fabrication compared to electroforming. However, \textcolor{black}{in general} these horns are fed by circular waveguides, which require circular-to-rectangular waveguide transitions for circuit integration. Therefore, the classical diagonal horn is widely utilized among feed horns due to its simplified manufacturing process and seamless integration with rectangular waveguides. Notably, it has symmetrical E- and H-planes, suitable for both circular and linear polarization. 

First introduced by Li in 1952 \cite{Tli1952}, diagonal horns feature a square aperture fed by a rectangular waveguide, with two transition sections converting the transverse-electric TE$_{10}$ mode to TE$_{11}$ in a circular waveguide and then to a diagonally polarized wave in the square waveguide - a superposition of two TE$_{10}$ and TE$_{01}$ waveguide modes. Subsequently, Love experimentally verified its radiation pattern and coined the term 'Diagonal horn' \cite{Love}. Later, Johansson and Whyborn showed that the intermediate circular section could be omitted and proposed E-plane split-block realization for submillimeter wavelengths \cite{Johansson_1992}. However, fabricating these components at high frequencies ($>$1 THz) with micron-level precision presents significant challenges. The performance of diagonal horns depends on the phase and amplitude balance between the TE$_{10}$ and TE$_{01}$ modes. Any asymmetry, such as E-plane misalignment, can lead to mode imbalance, increasing cross polarization levels, and result in an elliptically polarised beam. The influence of mode amplitude imbalance was discussed by Withington and Murphy in \cite{Murphy1992}, but the role of a phase between the modes has yet to be reported in the literature. Slight differences in propagation constants will cause a significant phase difference \cite{jorgarftg} at the antenna aperture. Previous studies have documented similar issues; Kerr \textit{et al.} noticed degradation in isolation for ortho-mode transducers (OMT) \cite{Kerr2009} and Ellison \textit{et al.} observed discrepancies in the radiation pattern of a 3.5-THz quantum-cascade laser (QCL) integrated with dual diagonal feed horns due to fabrication imperfections and misalignment \cite{Ellison2019}. Montofre \textit{et al.}, \cite{Montofre2019} noticed unbalanced cross polarization levels in the D-plane \textcolor{black}{diagonal-spline horn} and concluded that this is the effect of asymmetry between the blocks. Jayasankar \textit{et al.} also reported higher conversion losses in a 3.5-THz harmonic mixer with a diagonal horn in a WM64 waveguide and attributed it to high loss in the RF chain \cite{divya2021}.

In this paper, we study the performance degradation of THz diagonal horn antennas due to E-plane misalignment \textcolor{black}{since it is the most probable error due to alignment tolerances of split-block components}. The diagonal horn is a scaled version of the 4.7-THz antenna presented in \cite{divya202247} and the experiment is conducted at the WM570 frequency band to precisely control misalignment and to have a reliable antenna characterization. The rest of the paper is organized as follows: in section II, we have proposed an analytical model to compute the phase and amplitude imbalance caused by E-plane asymmetry in diagonal horns. We present the analytical results on the impact of misalignment on cross polarization loss at three frequencies, $k/k_c =$ 1.3, 1.5, and 1.8, along with electromagnetic (EM) simulations. Following this, we present the mechanical design that allows \textcolor{black}{controlled} misalignment and machining of the WM570 diagonal horn. In section III, we present the near-field measurement setup and experimental characterization of the horn. We compare three alignment scenarios, a) perfectly aligned, b) misaligned by 22~$\mu$m and c) misaligned by 40~$\mu$m. Finally, we conclude by summarizing our findings and proposing an alternative solution to address this issue.


\section{Method}
A standard WM570 diagonal horn (without the intermediate circular transition) was designed as described by Johansson and Whyborn \cite{Johansson_1992}, see Fig. ~\ref{3dem}. A square aperture with the side, $d=2.86$~mm, length $L=15$~mm, and corresponding flare angle $\theta=$ tan$^{-1}(d/\sqrt{2}L)=7.7^\circ$ were used, which results in a nominal antenna gain of circa 23 dBi. The feed is a standard rectangular waveguide (a $\times$ b) of dimension 570~$\times$~285 $\mu$m$^2$. To experimentally study the effect of E-plane misalignment $\delta$, guide structures were machined in a split block to allow for alignment and controlled misalignments \textcolor{black}{up to 40 $\mu$m corresponding to about 6\% of free-space wavelength.} This section describes the theory, EM simulations, mechanical design, machining and near-field characterization for analyzing the impact of \textit{E}-plane misalignment.

\subsection{Theoretical analysis}

\begin{figure}[t!]
    \centering
    \includegraphics[width = 1\linewidth]{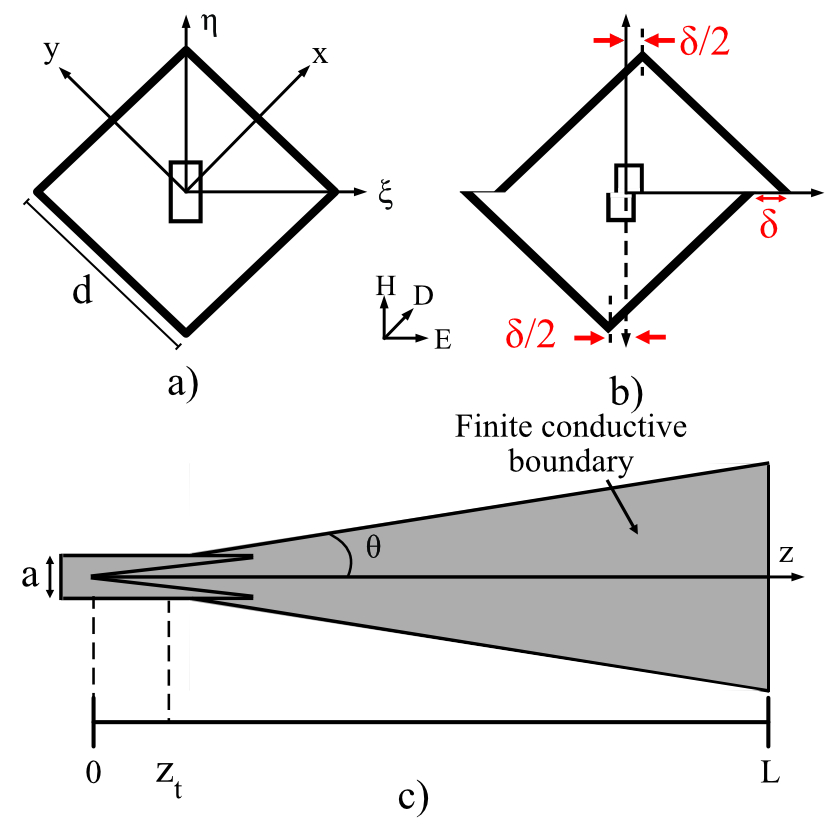}
    \caption{\textcolor{black}{a-b) Schematic of aligned and misaligned diagonal horn aperture. c) Diagonal horn with rectangular waveguide of width \textit{a}, length \textit{L} and flare angle $\theta$.}}
    \label{3dem}
\end{figure}

The aperture field distribution of a diagonal horn antenna \textcolor{black}{can be approximated as} an in-phase superposition of two orthogonal TE modes \cite{Johansson_1992}, TE$_{10}$ and TE$_{01}$, 

\begin{subequations}
\begin{gather}
\label{E_ap}
    \mathbf{E_{ap}} = \biggl [ E_{01} cos \bigg (\frac{\pi y}{d} \bigg )\hat{x} + E_{10} cos \bigg (\frac{\pi x}{d} \bigg) \hat{y} \biggr] \cdot e^{jk\rho}  \\
     k \rho = \frac{2\pi}{\lambda_{o}} \biggl [ \frac{d^2-2x^2-2y^2}{4L} \biggr ]
\end{gather}
\end{subequations}

\noindent where $\lambda_o$ is the free space wavelength and $k_\rho$ is the spherical phase front \cite{Muehldorf1970}. When the split blocks are aligned as shown in Fig.~\ref{3dem}a, the modes have phase and amplitude balance and exhibit the same \textcolor{black}{propagation phase constant}, $\beta$, which is not valid for the misaligned and distorted cross-section shown in Fig.~\ref{3dem}b. However, when the split blocks are misaligned, the original square cross-section of the \textcolor{black}{horn}, see Fig.~\ref{3dem}a, can be replaced with a rectangular shape, Fig.~\ref{3dem}b, where one side of the cross-section becomes effectively narrower $d-\delta/\sqrt{2}$ and the other broader $d+\delta/\sqrt{2}$. As a result, the two corresponding TE modes will have different \textcolor{black}{propagation phase constants} ($\beta^{+}$, $\beta^{-}$) and wave impedances, creating a phase and amplitude imbalance along the length of the diagonal horn. In general, the imbalance between the two modes at the aperture can be expressed as
\begin{align}
    \frac{E_{10}}{E_{01}} = \sqrt{\Omega} \cdot e^{j\varphi}
\end{align}
where $\Omega$ is power balance factor proposed by Withington and Murphy \cite{Murphy1992}, and $\varphi$ is the total phase imbalance at the aperture.
For a modest \textcolor{black}{and typical flare angle ($<10^\circ$)}, neglecting reflections, the total phase imbalance can be obtained as the sum of propagating infinitesimal waveguide sections \textit{dz}.

  \begin{align*}
  \label{phase}
    \varphi = \int_{z_t}^{L} \biggl ( \beta^{+}(z) - \beta^{-}(z) \biggr ) dz 
\end{align*}

\begin{align}
    & = \int_{z_t}^{L} \Biggl [ \sqrt{k^2-\biggl (\frac{\pi}{z \frac{d}{L}+\frac{\delta}{\sqrt{2}}} \biggr )^2} - \sqrt{k^2-\biggl (\frac{\pi}{z \frac{d}{L}-\frac{\delta}{\sqrt{2}}})^2} \Biggr ] dz
\end{align}

\begin{figure*}[b!]
    \centering
    \includegraphics[width=1\linewidth]{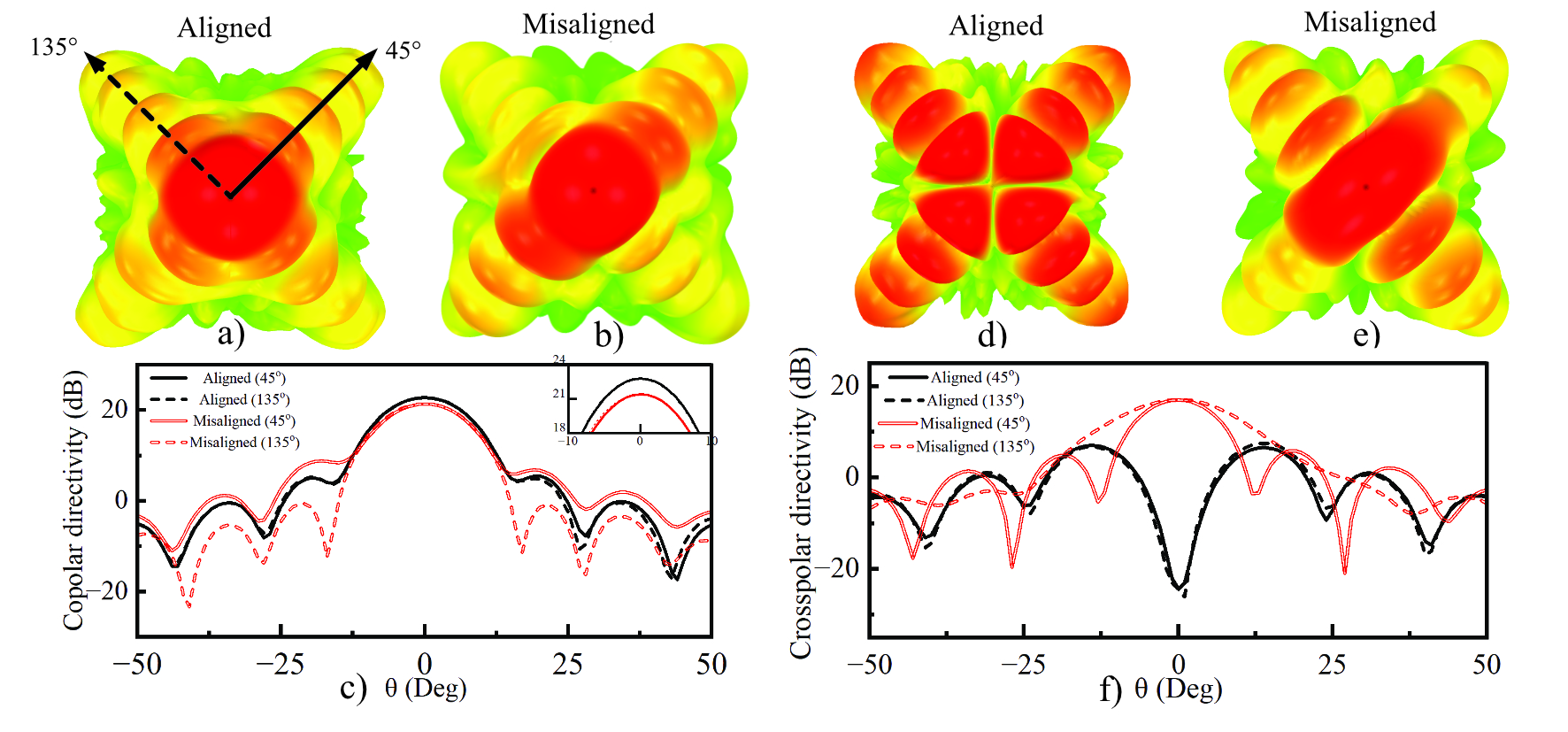}
    \caption{\textcolor{black}{Simulation of the diagonal horn's copolar and cross polar far-field radiation pattern at 470 GHz. (a-b) Copolar directivity of aligned and misaligned antenna (40~$\mu$m) (c-f) D-plane cuts of the copolar far-field pattern, solid line for ($\phi = 45^\circ$) and dashed line for ($\phi = 135^\circ$) cut for both aligned (black) and misaligned (red) cases. The inset highlights the reduction in copolar directivity of about 1.5~dB due to misalignment. (d-e) Cross polar directivity of aligned horn antenna with a null in boresight and misaligned antenna shows a central cross polarization component. }
    }
    \label{codir}
\end{figure*}
   
where $L$ is the length of the horn, $z_{t} = c \cdot L/d$ is the assumed starting point of the horn (throat), and \textit{c} is the corresponding cross-section width at the throat. Note that the wavenumber of each mode, $k_{c}$, changes with the position along the diagonal horn \textcolor{black}{\textit{z}-axis}, and is corrected with the misalignment factor $\pm \frac{\delta}{\sqrt{2}}$. The integral can be solved using analytical methods, leading to a lengthy expression. However, a first-order approximation can be found by introducing the parameter, $h=L \delta / (\sqrt{2} d)$, and reorganizing Eq. (\ref{phase}) as:

\begin{align*}
\varphi \cong 2h \int_{z_t}^{L} \biggl ( \frac{\beta(z+h) - \beta(z-h)}{2h} \biggr ) dz  
\end{align*}
\begin{align*}
& \cong 2h \int_{z_t}^{L} \biggl( \frac{\beta(z)}{dz} \biggr)  dz 
\end{align*}
\begin{align*}
&  = \frac{\sqrt{2}L \delta}{d} \biggl [\beta(z) \biggr ]_{z_t}^{L}
\end{align*}
\begin{align}
&  = \frac{\sqrt{2}L \delta}{d} \biggl [ \sqrt{k^2- \bigg (\frac{\pi}{d} \bigg )^2} - \sqrt{k^2- \bigg (\frac{\pi}{c} \bigg )^2} \biggr ]
\label{delta_phase}
\end{align}

Hence, the phase imbalance depends mainly on the flare angle, frequency, and the cross-section at the throat. For a large aperture $d\gg c$, the phase imbalance can be further approximated as $\frac{\delta}{c} \frac{\pi}{tan \theta}$. Note that $c \in (0,3a/ \sqrt{8})$ is a model fitting parameter representing the effective width of the throat cross-section.

Next, the amplitude or power imbalance results from uneven excitation of the two modes in the transition between the rectangular waveguide and the diagonal horn. For simplicity, we \textcolor{black}{neglect losses and} assume that the power imbalance only depends on the ratio of mode impedances at the throat. Using the power-voltage characteristic impedance definition (wave impedance scaled with waveguide height and width) \cite{Schelkunoff_1944, Williams2015}, the power balance, $\Omega$, can be estimated as:

    \begin{align*}
    \Omega = \frac{Z^{-}(z_{t})}{Z^{+}(z_{t})}= \left(\frac{c^{-}}{c^{+}}\right)^2 \cdot \frac{\beta^{-}(z_{t})}{\beta^{+}(z_{t})}  
\end{align*}

\begin{align}
    & = \left(\frac{c - \delta / \sqrt{2} }{c + \delta / \sqrt{2}} \right)^2 \cdot \frac{\beta^{-}(z_{t})}{\beta^{+}(z_{t})} 
    \label{omega}
\end{align}

\renewcommand{\thefigure}{4}
\begin{figure*}[hb!]
    \centering
    \includegraphics[width=1\linewidth]{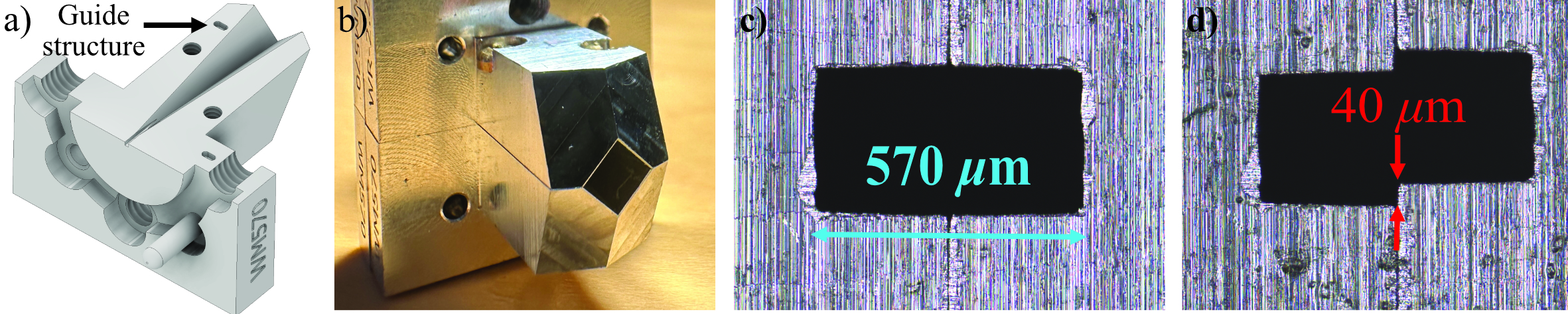}
    \caption{\textcolor{black}{WM570 diagonal horn antenna. a) CAD model of the antenna. b) Photograph of the machined horn antenna in Aluminum E-plane split-blocks. c) Micrograph of WM570 waveguide aperture when aligned. d) Micrograph of WM570 waveguide aperture with 40-$\mu$m misalignment.}}
    \label{cad}
\end{figure*}

Together with the aperture field, given in Eq.\ref{E_ap}), the analytical model (\ref{delta_phase}-\ref{omega})  provides insights and can be used to predict imbalance in phase and amplitude between the two modes, thereby allowing the calculation of the distorted aperture field for small misalignments ($\delta \ll \lambda$). The analytical model was validated with electromagnetic simulations of cross polarization coupling loss $\eta^-$:
\begin{align}
    \eta^- = \frac{P_{cr}}{P_{cr}+P_{co}},
\end{align}

where $P_{co}$ and $P_{cr}$ are the radiated copolar and cross polar power, respectively.  For a perfectly aligned antenna, 9.5\% of the total power is radiated into the cross polarized component as shown in Fig.~\ref{crosspolJoakim} and it agrees with the theoretical prediction in \cite{Murphy1992,Johansson_1992}. For the analytical model, a throat cross-section ($c/a$) of 0.82, 0.77, and 0.74 was used for the lower (1.3$f_c$), mid (1.5$f_c$) and upper (1.8$f_c$) frequency bands, respectively, which resulted in a good agreement with the EM simulations. Note: $f_c$ is the cutoff frequency of the waveguide.


\renewcommand{\thefigure}{3}
\begin{figure}[ht!]
    \centering
%
%
\definecolor{mycolor1}{rgb}{0.00000,0.44700,0.74100}%
\begin{tikzpicture}

\begin{axis}[%
width=7cm,
height=5cm,
at={(0.758in,0.481in)},
scale only axis,
xmin=0,
xmax=0.1,
xtick = {0,0.05,0.1},
xlabel style={font=\color{white!15!black}},
xticklabel style={
        /pgf/number format/fixed,
        /pgf/number format/precision=3
},
xlabel={$\delta/\lambda_o$},
ymin=-11,
ymax=0,
ytick = {-10,-5,0},
ylabel style={font=\color{white!15!black}},
ylabel={Cross-polar coupling loss $\eta$ (dB)},
ylabel style={align=center,text width=5cm},
axis background/.style={fill=white},
legend style={legend cell align=left, align=left, draw=white!15!black}
]
\node[black] at (axis cs:0.08,-7) {$f/f_c = 1.8$};
\node[black] at (axis cs:0.06,-2) {$f/f_c = 1.5$};
\node[black] at (axis cs:0.015,-5) {$f/f_c = 1.3$};
\addplot [color=black, dash dot, line width=2.0pt]
  table[row sep=crcr]{%
0	-10.2326824822945\\
0.0042	-10.056349436891\\
0.0084	-9.5683953156925\\
0.0126	-8.86529090695701\\
0.0168	-8.04929247415412\\
0.021	-7.19846205585065\\
0.0252	-6.3623357744115\\
0.0294	-5.56929835269273\\
0.0336	-4.83540733974686\\
0.0378	-4.17159426803943\\
0.042	-3.58955778195099\\
0.0462	-3.08001294480965\\
};

\addplot [color=black, dashed, line width=2.0pt]
    table[row sep=crcr]{%
0	-10.2326824822945\\
0.00483859649122807	-10.1179200260581\\
0.00967719298245614	-9.79181707630936\\
0.0145157894736842	-9.30054681403124\\
0.0193543859649123	-8.70022659043599\\
0.0241929824561403	-8.04166315571237\\
0.0290315789473684	-7.36348176190682\\
0.0338701754385965	-6.6918372299784\\
0.0387087719298246	-6.04300153567802\\
0.0435473684210526	-5.42638027762435\\
0.0483859649122807	-4.84698737015198\\
0.0532245614035088	-4.30721569168298\\
0.0580631578947368	-3.80803343344626\\
0.0629017543859649	-3.34978958437159\\
0.067740350877193	-2.93279079870787\\
0.072578947368421	-2.55778606096257\\
0.0774175438596491	-2.22649497969015\\
0.0822561403508772	-1.942368712417\\
0.0870947368421053	-1.71195218663404\\
0.0919333333333333	-1.54777445291248\\
};

\addplot [color=black, line width=2.0pt]
   table[row sep=crcr]{%
0	-10.2326824822945\\
0.0057719298245614	-10.1551918707958\\
0.0115438596491228	-9.93120503097294\\
0.0173157894736842	-9.58339247566725\\
0.0230877192982456	-9.14199735379676\\
0.028859649122807	-8.63795478844902\\
0.0346315789473684	-8.09827366901855\\
0.0404035087719298	-7.5441396254565\\
0.0461754385964912	-6.99093279860229\\
0.0519473684210526	-6.4491607307468\\
0.057719298245614	-5.92561899788204\\
0.0634912280701754	-5.42444264064674\\
0.0692631578947368	-4.94793845244959\\
0.0750350877192982	-4.49719801887553\\
0.0808070175438596	-4.07253066958076\\
0.086578947368421	-3.6737614925868\\
0.0923508771929825	-3.3004331946455\\
0.0981228070175439	-2.95194147069588\\
0.103894736842105	-2.62762530028003\\
0.109666666666667	-2.32682717801964\\
};

\addplot [mark=square*, only marks, mark options={solid, red}, mark size = 2.5 pt]
  table[row sep=crcr]{%
0	-10.42\\
0.0156739811912226	-9.57\\
0.0313479623824451	-8.28\\
0.0470219435736677	-6.77\\
0.0626959247648903	-5.42\\
0.0783699059561129	-4.27\\
0.0940438871473354	-3.38\\
0.109717868338558	-2.64\\
};

\addplot [mark=triangle*, only marks, mark options={solid, blue}, mark size = 3.5 pt]
table[row sep=crcr]{%
0	-10.4\\
0.0131333333333333	-9.48\\
0.0262666666666667	-7.6\\
0.0394	-5.87\\
0.0525333333333333	-4.37\\
0.0656666666666667	-3.28\\
0.0788	-2.5\\
0.0919333333333333	-1.98\\
};

\addplot [mark=diamond*, only marks, mark options={solid, green}, mark size = 3.5 pt]
table[row sep=crcr]{%
0	-10.3\\
0.0114	-8.8\\
0.0228	-6.7\\
0.0342	-4.82\\
0.0456	-3.46\\
};

\end{axis}

\end{tikzpicture}%
    \caption{\textcolor{black}{Cross-polar coupling loss. Increase in the fraction of power radiated into the cross polarized component due to E-plane misalignment for a diagonal horn antenna with a flare angle of 7.7$^\circ$ and an aperture size $d/a = 5$. The black lines represent the analytical model and the markers correspond to EM simulations. } }
    \label{crosspolJoakim}
\end{figure}

\subsection{EM simulation}

The EM simulation of the diagonal integrated with rectangular waveguide as shown in Fig.~\ref{3dem} was carried out using a commercial finite-element method based solver. A finite conductive boundary (aluminum) with electrical conductivity $\sigma = 3.8 \times 10^{7}$~S/m was assigned to the antenna and waveguide walls. A radiation boundary box was placed at the end of the horn to obtain the far-field responses. The side with the horn aperture was assigned a finite conductive boundary, and the rest of the faces were assigned radiation boundaries as mentioned in \cite{Hesler2001}. Two-fold symmetry can significantly reduce the computational time for the perfectly aligned case. However, symmetrical boundary conditions are inapplicable for misaligned configurations, necessitating the simulation of the entire horn and thus increasing the computational time. Simulated directivity is about 23 dB and E-plane \textcolor{black}{half-power beam width (HPBW)} is 13 degrees at 470 GHz. 
Fig. \ref{codir}a-b,d-e shows the simulated far-field radiation pattern of co and cross polar component of the diagonal horn at 470 GHz, respectively, for both aligned and misaligned antenna. Fig.~\ref{codir}c,f shows the D-plane cuts at $\phi = 45^\circ$ and 135$^\circ$. In Fig.~\ref{codir}f, we see a null in the cross polar component at boresight as expected for the aligned case \cite{Johansson_1992, Ludwig73} and when misaligned, shows an increased cross polar component at the boresight. Likewise, the copolar directivity also reduces by 1.5~dB when misaligned, as highlighted in the inset of  Fig.~\ref{codir}c.

\subsection{Mechanical design}

Fig.~\ref{cad}a shows the CAD model of the 470-GHz diagonal horn antenna integrated with a WM570 rectangular waveguide in an E-plane split-block housing. It has a waveguide flange based on UG-387 specification \cite{Waveguide}. \textcolor{black}{The length of the access waveguide to the horn was designed as short as possible, about one wavelength, to minimize attenuation.} A 45$^\circ$ chamfer was implemented \textcolor{black}{on the aperture plane} to redirect the reflected signal away from the optical axis. 

The top split block has trenches along the edge of the waveguide and horn to ensure a tight fit between the blocks while assembling. To facilitate accurate alignment, two rectangular guide structures were incorporated onto the top surface of the split blocks. The bottom block features guide holes measuring 1045 $\times$ 510~$\mu$m$^2$, while the top block employs pins sized at 1000 $\times$ 500~$\mu$m$^2$. The intentional allowance in tolerance permits controlled sliding of the blocks, thus enabling effortless alignment and misalignment of the split blocks. Micrographs of two extreme cases are shown in Fig.~\ref{cad}c-d. The misalignment was measured using a metrology microscope with $\pm$ 1 $\mu$m accuracy. Using guide structures offered advantages over traditional alignment methods, such as dowel pins, by minimizing rotational misalignments.

\subsection{Machining}
The diagonal horn was machined from a single block of aluminum (Al 6082) alloy using a CNC machine (KERN Micro HD) with micrometer precision. First, planarization of the top surface was carried out using an end mill of 250~$\mu$m-diameter. After machining the UG387 flange, the diagonal horn's aperture was machined using a 45$^\circ$ chamfer mill with a top radius of 5 $\mu$m, which resulted in a fillet with the same radius as the tool. Following that, a rectangular waveguide was machined using a 200-$\mu$m tungsten end mill with sharp corners. 

Burrs along the edge of the waveguide were removed since it could get folded into the waveguide cavity when the split-block halves are assembled. In addition, minor machining artifacts, especially at the horn's throat, were removed to avoid any reflections affecting the antenna's performance. After deburring, blocks were cleaned in an ultrasonic bath with acetone and isopropanol. The diagonal horn antenna machined in the E-plane split-block is shown in Fig.~\ref{cad}b. The cross-section of the WM570 waveguide aperture for the aligned and misaligned cases is shown in Fig.~\ref{cad}c-d.


\subsection{Measurement setup}

\renewcommand{\thefigure}{8}
\begin{figure*}[b!]
    \centering
    \input{Fig/cocross_full}
    \caption{\textcolor{black}{D-plane far-field radiation patterns. Comparison of simulation and measured copolar and cross polar radiation pattern of diagonal horn at 470 GHz in the D-plane ($\phi = 135^\circ$). a) perfectly aligned horn, b) misaligned by 22~$\mu$m and c) misaligned by 40~$\mu$m.} }
    \label{cocross}
\end{figure*}

Near-field pattern of the antenna was measured using a Keysight Vector Network Analyzer (VNA) N5242B and WM570 VDI frequency extenders as shown in the Fig.~\ref{setup}. The antenna under test (AUT) was connected to extender 1 on the left. For near-field scanning, a WM570 open-ended rectangular waveguide was connected to the right, which was mounted on a motorized linear XY stage with a step resolution of 2.5 $\mu$m. Input signals from direct digital synthesizers (DDS) in the VNA were fed to the cascaded Schottky multiplier chain in the frequency extenders to generate output signals in the range of \SIrange{325}{500}{GHz}. 

\renewcommand{\thefigure}{5}
\begin{figure}[H]
    \centering
\includegraphics[width=1\linewidth,keepaspectratio]{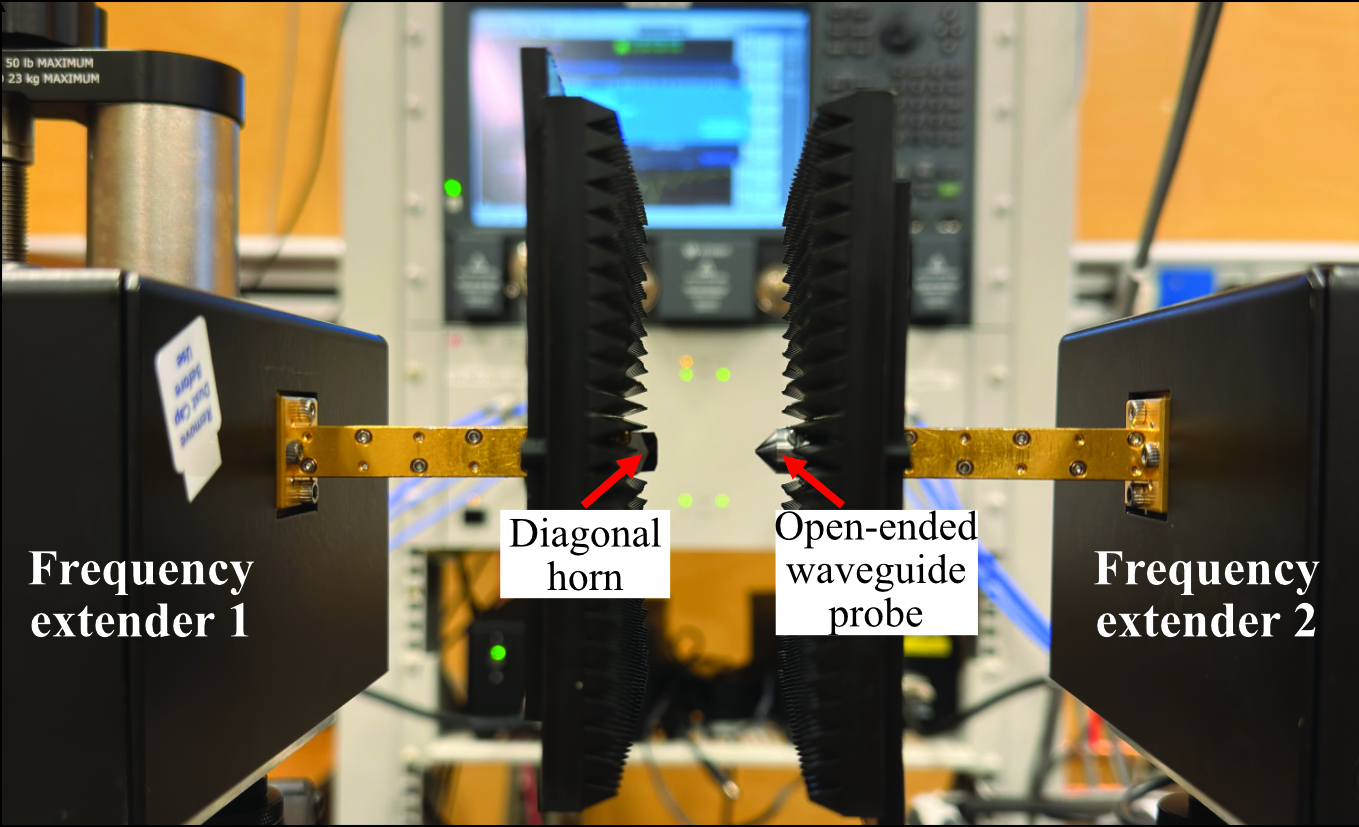}
    \caption{Photograph showing the near-field measurement setup with VDI's WM570 frequency extenders, diagonal horn antenna and an open-ended waveguide probe. }
    \label{setup}
\end{figure}

The amplitude and phase distribution of the diagonal horn antenna was measured at a distance of approximately \SI{1.5}{cm} \textcolor{black}{between horn's aperture and the probe}. The separation distance is a trade-off; while reducing the distance gives a smaller required scan area, care has to be taken to limit the standing waves and evanescent fields \cite{Balanis-2012-antenna}. A raster pattern was implemented to scan the rectangular XY grid of size 22 $\times$ 22~mm$^2$ with uniform spacing set to $\Delta x = \Delta y \leq \lambda/2$ \textcolor{black}{(300~$\mu$m)}. The setup was mounted on a vibration-free optical test bench from Thor Labs; absorbers were used to reduce the free-space standing waves in the optical axis, and the bench was layered with Eccosorb AN-73.

\section{Results}
\label{results}
First, the input return loss of the aligned horn antenna and WM570 open-ended waveguide probe was measured. Two-port short-offset short-load-thru (SOLT) calibration was carried out, and the IF bandwidth was set to 300 Hz. \textcolor{black}{Measured input return loss of the horn is better than 22~dB throughout the whole band from \SIrange{325}{500}{GHz} as shown in Fig.~\ref{s11}. For comparison, simulation results of the diagonal horn are also included which match closely to the measurement results.} 
\renewcommand{\thefigure}{6}
\begin{figure}[H]
    \centering
    \input{Fig/rlloss}
    \caption{\textcolor{black}{Return loss. Comparison of measured and simulated S$_{11}$ versus frequency of the aligned diagonal horn antenna and measured return loss of the WM570 open-ended rectangular waveguide probe.} }
    \label{s11}
\end{figure}

The probe was moved at a low-speed for near-field scanning to avoid smearing effects. The port cables for extender 1 with AUT are stationary. However, as the cable movement in extender 2 is unavoidable, a reference measurement was recorded at the boresight of the AUT after scanning each row, which was later used to make linear compensation of phase drift. An average magnitude drift of about $\pm$ 0.01 dB and phase drift of $\pm$ 10 degrees were recorded during the complete scan. The probe was aligned to the boresight of the AUT by employing a phase-detection scheme that finds the local phase minima of S$_{21}$. The magnitude and phase of the tangential electric field components of the antenna were measured, and standard near-field to far-field transformation technique (\textit{NF/FF}) \cite{Yaghjian1986, Gibson2017} was employed to compute the far-field radiation pattern of the antenna. For cross polar measurements, a 90$^\circ$ WM570 VDI waveguide twist was used. 

\renewcommand{\thefigure}{7}
\begin{figure}[t!]
    \centering
    \input{Fig/eh}
    \caption{\textcolor{black}{E- and H-plane far-field radiation patterns. Comparison of simulation and measured copolar radiation pattern of aligned diagonal horn antenna at 470 GHz in two principal planes: a) E ($\phi = 0$), and b) H ($\phi = 90$).}}
    \label{EH}
\end{figure}

Three cases were studied in detail: i) perfectly aligned, ii) misaligned by $\delta = 0.035\lambda_o$ (22~$\mu$m), and iii) misaligned by $\delta = 0.064\lambda_o$ (40~$\mu$m) where $\lambda_o$ is the free space wavelength at 470~GHz. The far-field cuts of an aligned diagonal horn antenna in E ($\phi$ = 0$^\circ$) and H ($\phi$ = 90$^\circ$) are shown in Fig.~\ref{EH}. The directivity of the perfectly aligned diagonal antenna is measured at 22.8 dB which is in good agreement with the simulation. As expected, the E- and H-planes are symmetrical, with equal beam widths and high sidelobe levels \textcolor{black}{about 15~dB} in the D-plane, refer Fig.~\ref{cocross}a.

The far-field radiation pattern of aligned and misaligned antennas is shown in Fig.~\ref{cocross}. When misaligned by 40~$\mu$m, the copolar directivity reduces by 1.5~dB at 470 GHz as shown in Fig.~\ref{cocross}c. The mode imbalance due to the E-plane misalignment results in a central cross-polarized lobe \cite{Murphy1992}. \textcolor{black}{A 2D-contour of co- and cross polar radiation patterns  of both aligned and misaligned antenna ($40~\mu$m) is shown in Fig.~\ref{contaligned}.} \textcolor{black}{A short planar scanning range was chosen to reduce the measurement time, which restricts the validity to a limited angular span \cite{Bieren2014}. }


\renewcommand{\thefigure}{9}
\begin{figure*}
    \centering
    \includegraphics[width=0.8\linewidth]{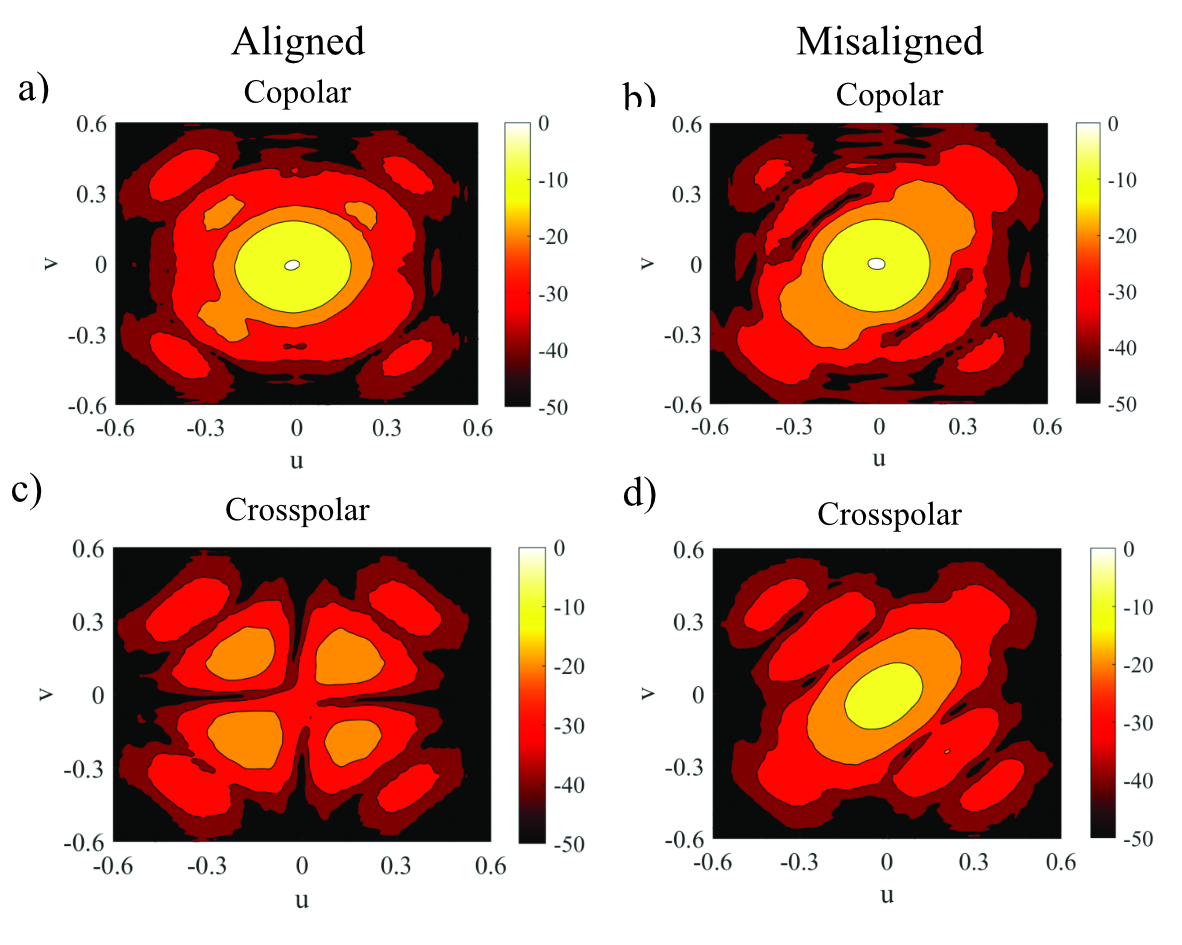}
    \caption{\textcolor{black}{Measured radiation pattern of aligned (left) and misaligned antenna (right) at 470 GHz. a-b) co-polarization, c-d) cross-polarization. Normalized to peak copolar directivity of 23~dB.}}
    \label{contaligned}
\end{figure*}

\section{Discussion}
The optical coupling to a Gaussian beam (Gaussicity) is an important parameter for quasi-optical system design. \textcolor{black}{It is defined as the maximum power coupling of the beam produced by the horn to a linearly polarized Gaussian beam} and it can be calculated from aperture E-fields as shown below \cite{gold, Johansson95}, 
\begin{align}
    \eta_{gauss} = \frac{\Big| \bigl \langle \hat{E_{ap}}~| ~\hat{g}\bigl \rangle \Big|^2}{\bigl \langle \hat{E_{ap}}~| ~\hat{E_{ap}}\bigl \rangle \bigl \langle \hat{g}~| ~\hat{g}\bigl \rangle}
\end{align}

\renewcommand{\thefigure}{10}
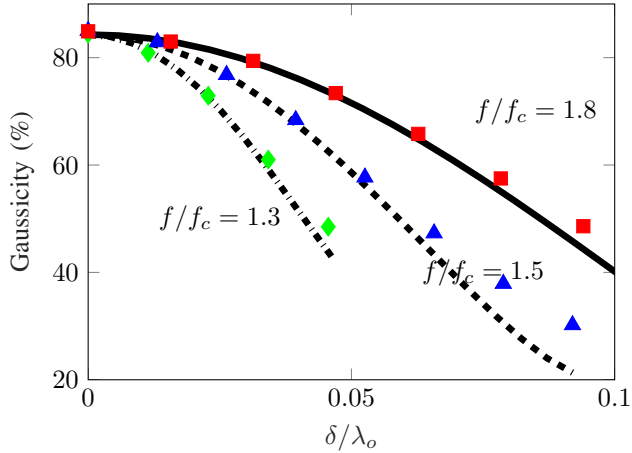
\begin{figure}[t!]
    \centering
    \begin{tikzpicture}

\begin{axis}[%
width=7cm,
height=5cm,
at={(0.758in,0.481in)},
scale only axis,
xmin=0,
xmax=0.1,
xtick = {0,0.05,0.1},
xlabel style={font=\color{white!15!black}},
xticklabel style={
        /pgf/number format/fixed,
        /pgf/number format/precision=3
},
xlabel={$\delta/\lambda_o$},
ymin=20,
ymax=90,
ylabel style={font=\color{white!15!black}},
ylabel={Gaussicity (\%)},
axis background/.style={fill=white},
legend style={legend cell align=left, align=left, draw=white!15!black}
]
\node[black] at (axis cs:0.085,70) {$f/f_c = 1.8$};
\node[black] at (axis cs:0.075,40) {$f/f_c = 1.5$};
\node[black] at (axis cs:0.025,50) {$f/f_c = 1.3$};
\addplot [mark=diamond*, only marks, mark options={solid, green}, mark size = 3.5 pt]
  table[row sep=crcr]{%
0	84.5\\
0.0114025085518814	80.9\\
0.0228050171037628	72.9\\
0.0342075256556442	61\\
0.0456100342075257	48.5\\
};

\addplot [mark=triangle*, only marks, mark options={solid, blue}, mark size = 3.5 pt]
  table[row sep=crcr]{%
0	85\\
0.0131406044678055	82.9\\
0.026281208935611	76.8\\
0.0394218134034166	68.4\\
0.0525624178712221	57.7\\
0.0657030223390276	47.3\\
0.0788436268068331	37.9\\
0.0919842312746386	30.2\\
};

\addplot [mark=square*, only marks, mark options={solid, red}, mark size = 2.5 pt]
  table[row sep=crcr]{%
0	84.9\\
0.0156739811912226	83\\
0.0313479623824451	79.4\\
0.0470219435736677	73.4\\
0.0626959247648903	65.8\\
0.0783699059561129	57.5\\
0.0940438871473354	48.6\\
0.109717868338558	39.8\\
};
\addplot [color=black, line width=2.5pt]
  table[row sep=crcr]{%
0	84.3024791653356\\
0.00577462464939779	84.1249785682013\\
0.0115492492987956	83.5937401474258\\
0.0173238739481934	82.7125490034151\\
0.0230984985975912	81.4876969623545\\
0.0288731232469889	79.927957671153\\
0.0346477478963867	78.0445520572455\\
0.0404223725457845	75.8511036891479\\
0.0461969971951823	73.3635860248869\\
0.0519716218445801	70.6002596118997\\
0.0577462464939779	67.5816013776286\\
0.0635208711433757	64.3302258195952\\
0.0692954957927735	60.8707987977283\\
0.0750701204421712	57.2299446194972\\
0.080844745091569	53.436147232182\\
0.0866193697409668	49.5196464898795\\
0.0923939943903646	45.5123306580387\\
0.0981686190397624	41.4476265731299\\
0.10394324368916	37.3603892148545\\
0.109717868338558	33.2867929098897\\
};

\addplot [color=black, dashed, line width=2.5pt]
  table[row sep=crcr]{%
0	84.3024791653356\\
0.00484127533024414	84.0384687671504\\
0.00968255066048828	83.2490479384096\\
0.0145238259907324	81.9420572478638\\
0.0193651013209766	80.1305974821386\\
0.0242063766512207	77.8330853451302\\
0.0290476519814648	75.0733433377475\\
0.033888927311709	71.8807402357867\\
0.0387302026419531	68.2904024655953\\
0.0435714779721972	64.343532508423\\
0.0484127533024414	60.0878864324436\\
0.0532540286326855	55.5784930864308\\
0.0580953039629297	50.8787467816804\\
0.0629365792931738	46.0620908294704\\
0.0677778546234179	41.214663848945\\
0.0726191299536621	36.4395744018084\\
0.0774604052839062	31.8640626659443\\
0.0823016806141504	27.65210170789\\
0.0871429559443945	24.0281099430476\\
0.0919842312746386	21.3260656012657\\
};

\addplot [color=black, dash dot, line width=2.5pt]
  table[row sep=crcr]{%
0	84.3024791653356\\
0.00420092420332473	83.8939240710868\\
0.00840184840664946	82.6729432264637\\
0.0126027726099742	80.6538871382085\\
0.0168036968132989	77.8617453839714\\
0.0210046210166237	74.3340727494745\\
0.0252055452199484	70.1244692273636\\
0.0294064694232731	65.3088307488487\\
0.0336073936265978	59.9969100362619\\
0.0378083178299226	54.3545981713597\\
0.0420092420332473	48.6456478687981\\
0.046210166236572	42.979451448611\\
};

\end{axis}

\end{tikzpicture}%
    \caption{\textcolor{black}{Coupling to a Gaussian beam. Degradation of Gaussicity due to E-plane misalignment for a diagonal horn antenna. For the analytical model, the same throat cross-section (c/a) in the range of 0.7-0.8 was used similar to Fig.~\ref{crosspolJoakim} for three frequency bands. The black lines represent the analytical model and the markers are data points from the corresponding EM simulation.} }
    \label{gauss}
\end{figure}

where $\hat{g}$ is the first-order Gaussian beam function.Fig.~\ref{gauss} shows the estimated performance degradation of the diagonal horn antenna due to E-plane misalignment along with the analytical model described earlier in Section IIA. The beam waist was kept constant as $w =$~0.86$\cdot d/2$ \cite{Johansson_1992} and note that the analytical model is valid only for small misalignment ($\delta$). Evidently, Gaussicity follows the same trend as the cross polarization loss in Fig.~\ref{crosspolJoakim} and is more sensitive when operating close to the cut-off frequency of the waveguide feed. We can observe 3-dB degradation in Gaussicity at 8\% misalignment relative to the wavelength when operating in the middle of the waveguide band. The degradation is even more severe in the low-frequency part of the waveguide band due to a larger dispersion ($\beta-k$), resulting in a larger imbalance between the two modes. Hence, operating in the upper part of the waveguide band \textcolor{black}{as shown in Fig.~\ref{gauss}} and using a larger flare angle, $\theta$, will reduce the influence of E-plane misalignment. Hence, there will be a trade-off between flare angle for optimum gain and robustness against fabrication and assembly tolerances. A possible \textcolor{black}{alternative} at very short wavelengths is to use a standard pyramidal horn, which is more or less immune to E-plane misalignment, thanks to the 'single' mode (TE$_{10}$) operation. Alternatively, use a horn with a throat that expands fast, such as the smooth-walled spline-profile diagonal horn \cite{Gibson2013}, or similar conical spline-horns \cite{hammar2016, granet2004}.


\section{Conclusion}

We have conducted a comprehensive study on the performance degradation of THz diagonal horn antennas due to E-plane misalignment. The consequences of misalignment have been quantified through a combination of theoretical analysis, EM simulations, and near-field antenna pattern measurements in the 325-500 GHz frequency range. The finite-element modeling matched very well with the proposed analytical model and experimental results, indicating that misalignment causes amplitude and phase imbalance in the TE$_{01}$ and TE$_{10}$ modes leading to high cross polarization loss. Particularly, when operating close to the cut-off frequency, $k_{c}$, of the waveguide feed, the diagonal horn is severely impacted by misalignment. \textcolor{black}{The $\delta/\lambda_o$ normalization in Fig.\ref{crosspolJoakim} and \ref{gauss} allows our findings to be easily scalable and applicable across various frequency bands.} These results offer valuable practical design guidelines that are critical,  especially beyond 3~THz. For future work, tolerance analysis work can be extended by addressing other critical horn parameters and comparing with other type of common terahertz horns.


\section*{Acknowledgement}
The authors would like to thank Mats Myremark for the high-precision machining of the diagonal horn antenna. The authors would also like to acknowledge the Kollberg Laboratory and thank Helena Rodilla and Patrik Blomberg for proofreading the manuscript.  

\bibliographystyle{IEEEtran}
\bibliography{IEEEfull,bibl}

 \begin{IEEEbiography}[{\includegraphics[width=1in,height=1.5in,clip,keepaspectratio]{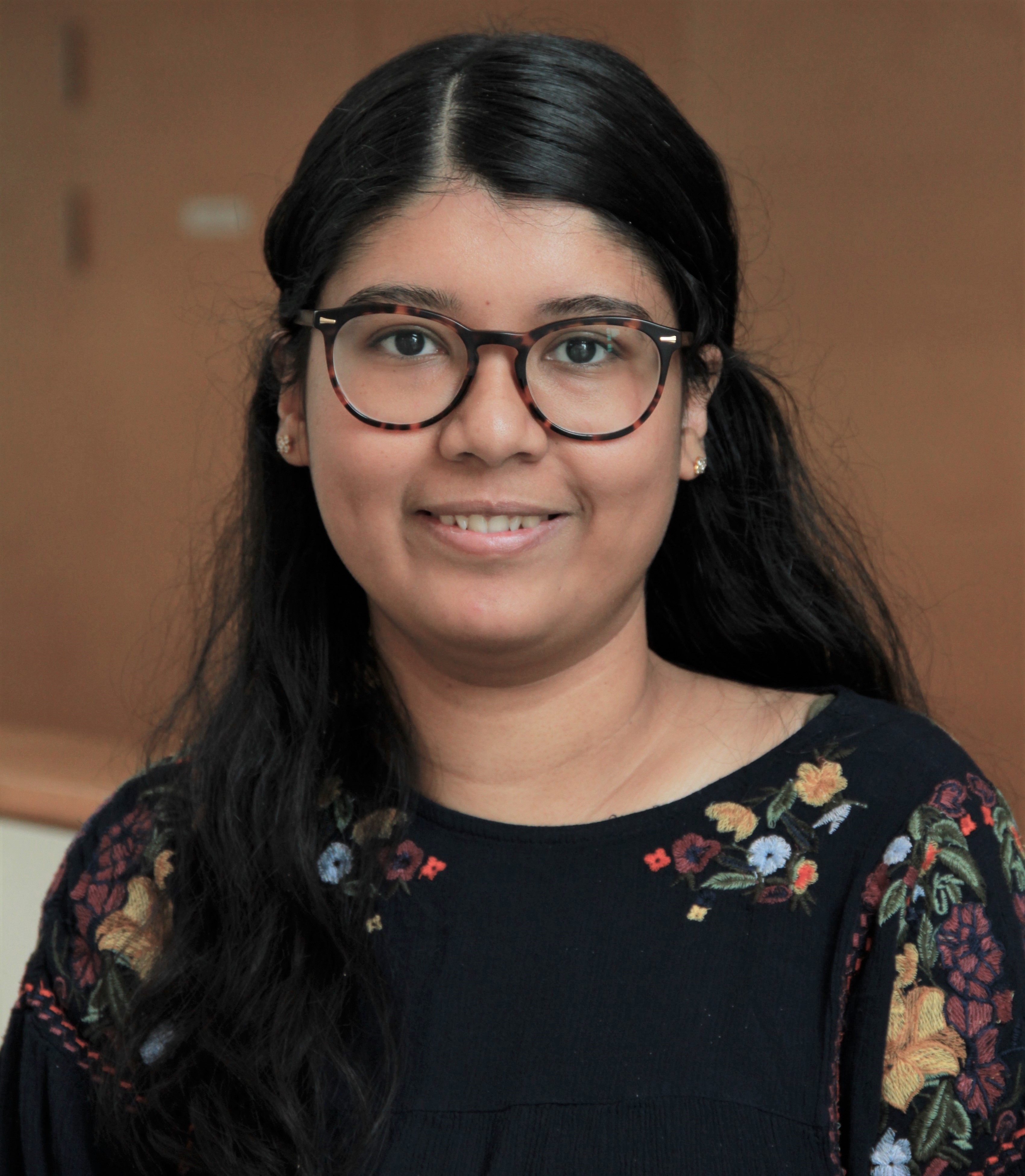}}]%
{Divya Jayasankar} (S’19) was born in India in 1994. She received her bachelor's degree in electronics and communication engineering in 2015. Followed by a master's degree in wireless, photonics and space engineering from Chalmers University of Technology, Sweden, in 2019.

From 2015-2017, she worked as a research assistant at Raman Research Institute in India. Since 2019, she has been employed as a doctoral student at the Research Institutes of Sweden (RISE) and the Terahertz and Millimeter-wave Laboratory at Chalmers University of Technology. Her research interests include the development of THz Schottky mixers and THz metrology. During her PhD, she worked as a visiting PhD student at the German Aerospace Center (DLR) in 2021, the University of Warwick in 2022, and Virginia Diodes Inc. (VDI) in 2023.

D. Jayasankar received the Ericsson Research Foundation award, the European Microwave Association (EuMA) internship award, the IEEE-MTTs graduate fellowship, and the Roger Pollard fellowship for microwave measurements by ARFTG. She also won second place in the best student paper competition at the ISSTT conference held in Baeza, Spain, in 2022.  
\end{IEEEbiography}

\begin{IEEEbiography}
[{\includegraphics[width=1in,height=1.5in,clip,keepaspectratio]{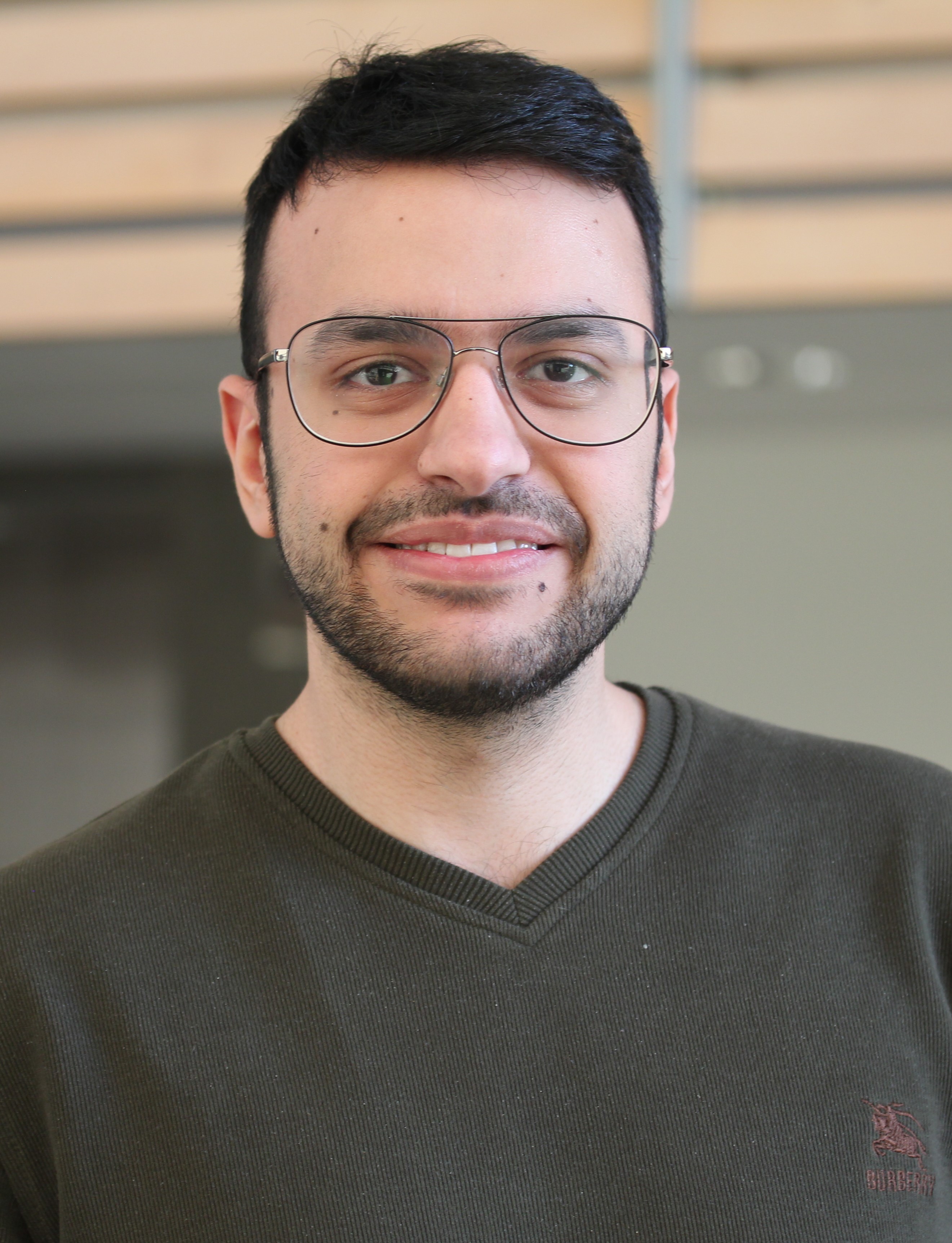}}]%
    {Andre Koj} received a bachelor's and master's degree in Engineering Physics from Chalmers University of Technology, in 2021 and 2023, respectively.
    
    He is currently a research engineer in the Quantum Technology department at Chalmers University Technology, Gothenburg, Sweden.
    
\end{IEEEbiography}

\begin{IEEEbiography}
[{\includegraphics[width=1in,height=1.5in,clip,keepaspectratio]{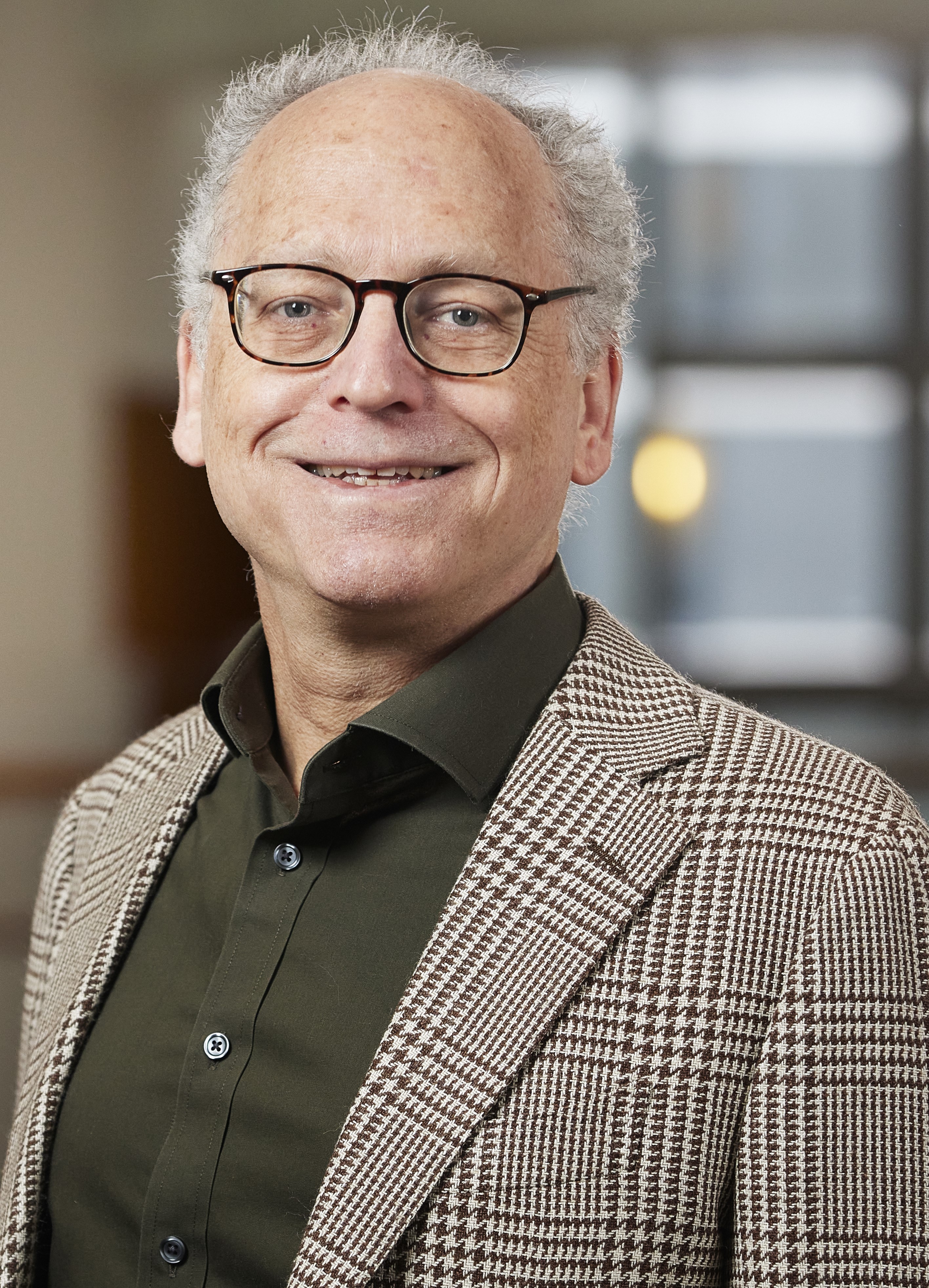}}]%
{Jeffrey Hesler} received the B.S.E.E. degree from Virginia Tech, Blacksburg, VA, USA, in 1989, and the M.S.E.E. and Ph.D. degrees from the University of Virginia, Charlottesville, VA, USA, in 1991 and 1996, respectively.

He is the president and Chief Technology Officer of Virginia Diodes Inc., Charlottesville, VA, USA, and a Visiting Research Assistant Professor at the University of Virginia. His career is focused on the creation of new technologies that are making possible the full exploitation of the Terahertz (THz) frequency band for scientific, defence, and industrial applications. He has authored and coauthored more than 150 technical papers in refereed international conferences and journals, given talks at THz-focused workshops and conferences such as IMS and EuMW. THz systems based on his innovative designs are now used in hundreds of research laboratories throughout the world. Dr. Hesler is a member of IEEE TC MTT-4 on THz Technology and Applications and serves as a reviewer for a variety of the IEEE and IEE journals.
In 2024, he became an Adjunct Professor at the Terahertz and Millimeter-wave Laboratory, Department of Microtechnology and Nanoscience (MC2), Chalmers University of Technology. 
\end{IEEEbiography}

\begin{IEEEbiography}[{\includegraphics[width=1in,height=1.5in,clip,keepaspectratio]{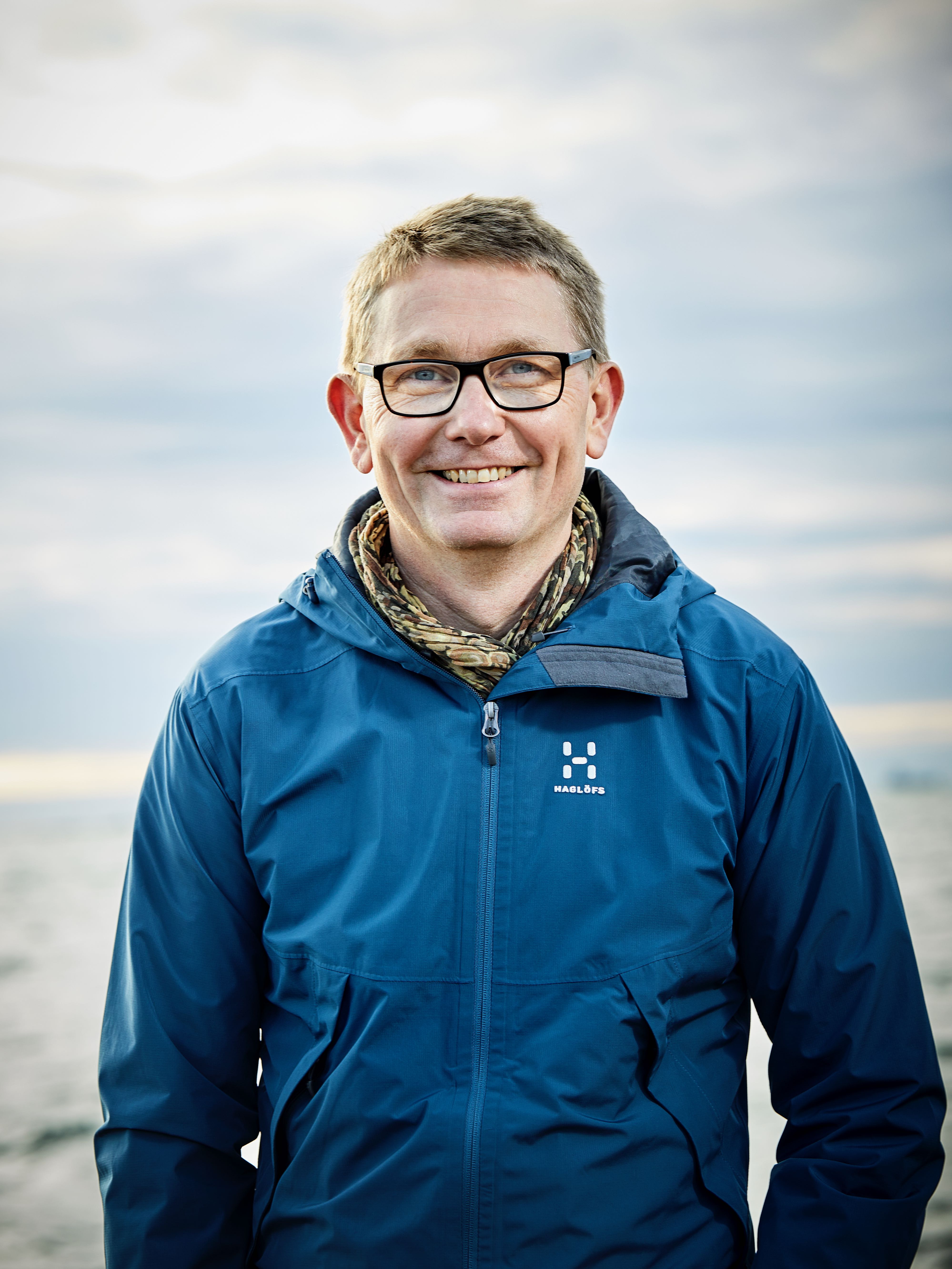}}]%
{Jan Stake}
(S’95–M’00–SM’06) was born in Uddevalla, Sweden, in 1971. He received an M.Sc. in electrical engineering and a Ph.D. in microwave electronics from Chalmers University of Technology in Gothenburg, Sweden, in 1994 and 1999, respectively.

In 1997, he was a Research Assistant at the University of Virginia, Charlottesville, VA, USA. From 1999 to 2001, he was a Research Fellow with the Millimetre Wave Group at the Rutherford Appleton Laboratory, Didcot, UK. He then joined Saab Combitech Systems AB, Gothenburg, Sweden, as a Senior RF/microwave Engineer until 2003. From 2000 to 2006, he held different academic positions with the Chalmers University of Technology, and from 2003 to 2006, he was also the Head of the Nanofabrication Laboratory, Department of Microtechnology and Nanoscience (MC2). In 2006, he was appointed Professor and the Head of the Terahertz and Millimetre Wave Laboratory at the Chalmers University of Technology. He was a Visiting Professor with the Submillimeter Wave Advanced Technology (SWAT) Group at Caltech/JPL, Pasadena, CA, USA, in 2007 and at TU Delft, the Netherlands, in 2020. He is also the co-founder of Wasa Millimeter Wave AB, Gothenburg, Sweden. His research interests include high-frequency semiconductor devices, terahertz electronics, submillimeter wave measurement techniques, and terahertz systems.

Prof. Stake served as the Editor-in-Chief for the IEEE Transactions on Terahertz Science and Technology between 2016 and 2018 and as Topical Editor between 2012 and 2015. From 2019 to 2021, he was chairperson of the IEEE THz Science and Technology Best Paper Award committee. He is an elected member of the International Society of Infrared, Millimeter and Terahertz Waves (IRMMW-THz) board.
\end{IEEEbiography}

\end{document}